\documentclass[preprint,
 amsmath,amssymb,
 aps,
]{revtex4-1}

\usepackage{graphicx}% Include figure files
\usepackage{dcolumn}% Align table columns on decimal point
\usepackage{bm}% bold math
\usepackage{hyperref}% add hypertext capabilities
\usepackage{color}      % use if color is used in text
\begin{document}

%\preprint{APS/123-QED}

\title{Radiative Decay of Neutron-Unbound Intruder States in $^{19}$O}
%\thanks{A footnote to the article title}%

 \author{R.Dungan}
% \altaffiliation[Also at ]{Physics Department, XYZ University.}%Lines break automatically or can be forced with \\
 \author{S.L. Tabor}
 \author{Vandana Tripathi}
 \author{A.Volya}
 \author{K.Kravvaris}
 \author{B.Abromeit}
 \author{D.D.Caussyn}
 \author{S.Morrow}
 \author{J.J. Parker IV}
 \author{P.-L. Tai}
 \author{J.M. VonMoss}

\affiliation{
 Department of Physics,Florida State University,Tallahassee,FL\\
 32306,USA
 %This line break forced with \textbackslash\textbackslash
}%

%\collaboration{MUSO Collaboration}%\noaffiliation

%\author{Charlie Author}
 %\homepage{http://www.Second.institution.edu/~Charlie.Author}
%\affiliation{
 %Second institution and/or address\\
 %This line break forced% with \\
%}%
%\affiliation{
 %Third institution, the second for Charlie Author
%}%
%\author{Delta Author}
%\affiliation{%
 %Authors' institution and/or address\\
 %This line break forced with \textbackslash\textbackslash
%}%

%\collaboration{CLEO Collaboration}%\noaffiliation

\date{\today}% It is always \today, today,
             %  but any date may be explicitly specified

\begin{abstract}
The $^{9}$Be($^{14}$C, $\alpha$$\gamma$) reaction at E$_{Lab}$=30 and 35 MeV was used to study excited states of $^{19}$O.  The Florida State University (FSU) $\gamma$ detector array was used to detect $\gamma$ radiation in coincidence with charged particles detected and identified with a silicon $\Delta$E-E particle telescope.  Gamma decays have been observed for the first time from six states ranging from 368 to 2147 keV above the neutron separation energy (S$_{n}$=3962 keV) in $^{19}$O.  The $\gamma$ decaying states are interspersed among states previously observed to decay by neutron emission.  The ability of electromagnetic decay to compete successfully with neutron decay is explained in terms of neutron angular momentum barriers and small spectroscopic factors implying higher spin and complex structure for these intruder states.  These results illustrate the need for complementary experimental approaches to best illuminate the complete nuclear structure.
%\begin{description}
%\item[Usage]
%Secondary publications and information retrieval purposes.
%\item[PACS numbers]
%May be entered using the \verb+\pacs{#1}+ command.
%\item[Structure]
%You may use the \texttt{description} environment to structure your abstract;
%use the optional argument of the \verb+\item+ command to give the category of each item. 
%\end{description}
\end{abstract}

%\pacs{Valid PACS appear here}% PACS, the Physics and Astronomy
                             % Classification Scheme.
%\keywords{Suggested keywords}%Use showkeys class option if keyword
                              %display desired
\maketitle
%\section{Introduction}
%\indent 

The decay mode(s) of a nuclear quantum state is (are) one of its most important properties after energy.  Usually nuclear decay follows the hierarchy of the fundamental forces of nature.  Decay by emission of one or more particles mediated by the strong nuclear interaction is normally the dominant mode for those states unbound to particle emission.  Bound excited states usually decay by emission of electromagnetic radiation to the ground state. The ground state, in turn, decays much more slowly by $\beta$ decay mediated by the weak nuclear interaction until the lowest energy neutron-to-proton ratio has been reached.  Of course, there are exceptions: lower energy charged particle decay from unbound states is inhibited by the Coulomb barrier, and large spin change or low emission energy can inhibit $\gamma$ decay so much that $\beta$ decay occurs first.  Even decay by neutrons which face no Coulomb barrier can be inhibited by large spin change, as is well known in high-spin spectroscopy of medium and heavy nuclei.  However, based on the simple picture of barrier penetrability, neutron decay is usually assumed to dominate over radiative decay when angular momentum barriers are not very high.  But Fermi's Golden Rule \cite{fermi} states that the decay rate is the product of the coupling strength, the density of final states, and the matrix element between the wavefunction of the parent state and the daughter state (usually called spectroscopic factor ($S$)). What is often overlooked is that this latter factor due to the nuclear structure may be more instrumental than the angular momentum barrier in limiting neutron decay rates below those of electromagnetic decay.  Examples are states with complex structure often involving intruder configurations.

\begin{figure}[!h]
\centerline{\includegraphics[scale=0.7]{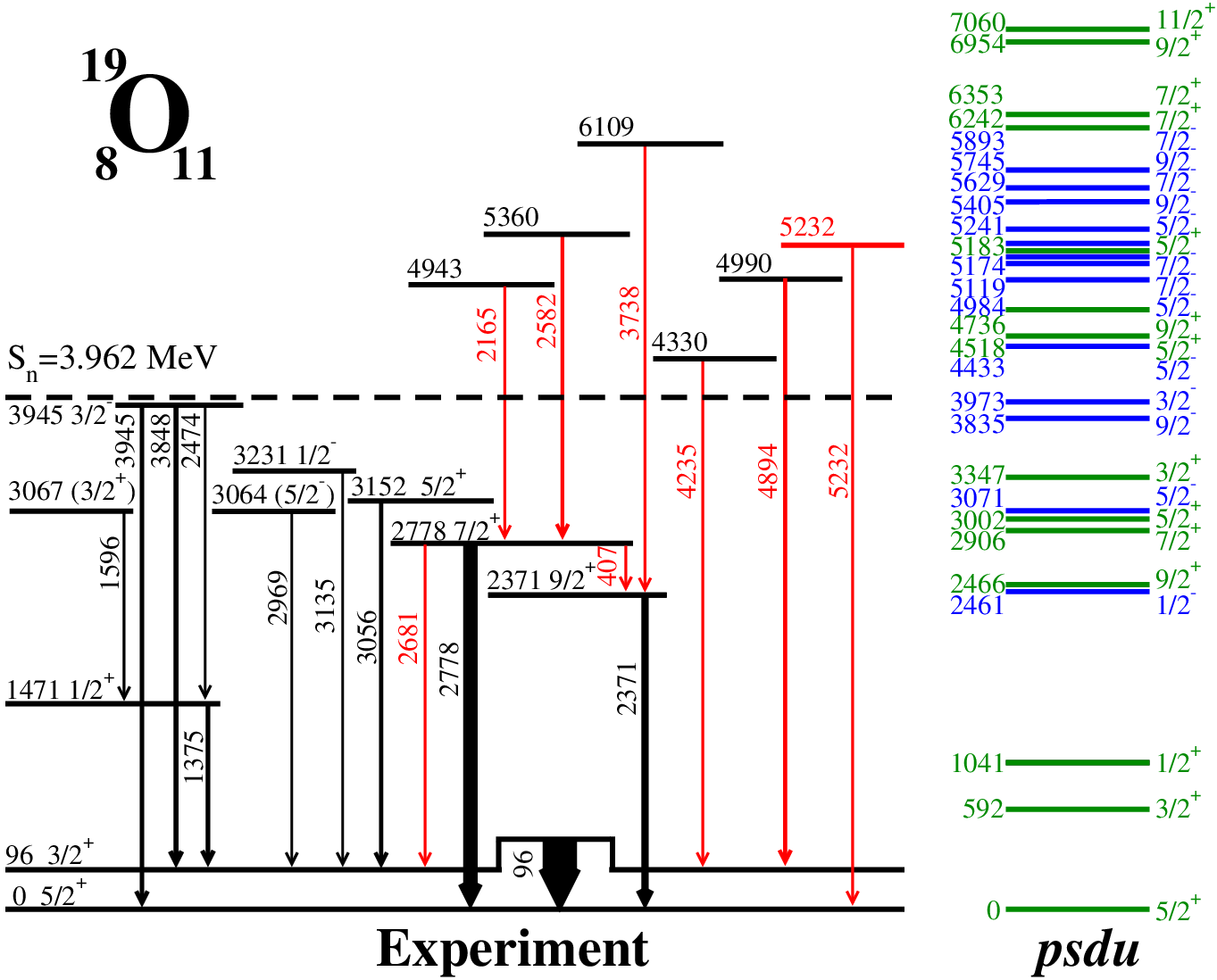}}
\caption{\label{fig:levelscheme} (Color online) The level and decay scheme of $^{19}$O along with shell model calculations using the \emph{psdu} interaction. Previously (Newly) observed states and $\gamma$ decays are colored in black (red) on the experimental level scheme. Positive (negative) parity levels are colored green (blue) for the calculated states using the \emph{psdu} interaction. All the calculated states predicted from the \emph{psdu} interaction are shown up to 4 MeV and all states with J $\geq$ 7/2 above 4 MeV are shown}
\end{figure}

The role of factors beyond barrier penetrability is highlighted by observations reported in this paper of the decays of unbound states in $^{19}$O populated in the $^9$Be($^{14}$C,$\alpha \gamma$) reaction which favors higher-spin states, which are typically more complex and may involve intruder configurations.  The importance of $^{19}$O with a closed major proton shell and an exactly half-filled neutron d$_{5/2}$ orbital has been recognized with over half a century of investigations~\cite{ahnlund54,zimmermann59,armstrong61,moreh64I,allen65,broude71,hibou71,crozier72,sen74,fortune74, fortune77,koehler,dufour86,warburton88,sumithrarachchi06,nagai07,oertzen10}.  In spite of this record, we were surprised to discover that six states from 368 to 2147 keV above the neutron threshold decay predominantly or totally by $\gamma$ radiation (Figure \ref{fig:levelscheme}). Five of these states had been previously reported~\cite{fortune77,oertzen10} in $^{13}$C($^7$Li,$p$) reaction studies which also favor higher-spin states, but no $\gamma$ detection was used in those works.  Many neutron decaying states in $^{19}$O were observed in an $^{18}$O + neutron resonance experiment~\cite{koehler}.  There is no evidence above statistical uncertainties for the $\gamma$ decaying states reported here in the neutron resonance spectrum~\cite{koehler}, implying $\Gamma_n$ widths much less than a keV.  The $\gamma$ decaying states lie interspersed among the neutron decaying ones, leading to one of the clearest pictures of the competition between neutron and radiative decay in light nuclei and emphasizing the need for different experimental approaches to gain a clear picture of nuclear structure, especially more exotic intruder states. 

Barrier penetrability alone cannot explain the difference between neutron and $\gamma$ decaying states since the spins of the $\gamma$ decaying ones cannot be very high due to their decays to known states in $^{19}$O with spins of 3/2$^+$ to 9/2$^+$.  They must have significantly smaller $S$ values for neutron decay, presumably due to more complex structure.

The $\gamma$ decaying unbound states were discovered following production of $^{19}$O in the $^{9}$Be($^{14}$C,$\alpha\gamma\gamma$) reaction performed at the Florida State University (FSU) John D. Fox Superconducting Laboratory using $^{14}$C beams of 30 and 35 MeV. The $^9$Be target thickness of 0.125 mm was chosen to allow $\emph{$\alpha$}$ particles to escape from the target with limited energy loss while stopping the $^{14}$C beam.  The target was placed in the FSU Compton-suppressed $\gamma$ detector array.  A Si $\Delta$E - E telescope was placed at 0$^{\circ}$ relative to the beam to detect and identify charged reaction products for channel selection. More experimental information can be found in Ref. \cite{vonmoss}.  

\begin{figure}[!h]
\centerline{\includegraphics[scale=0.7]{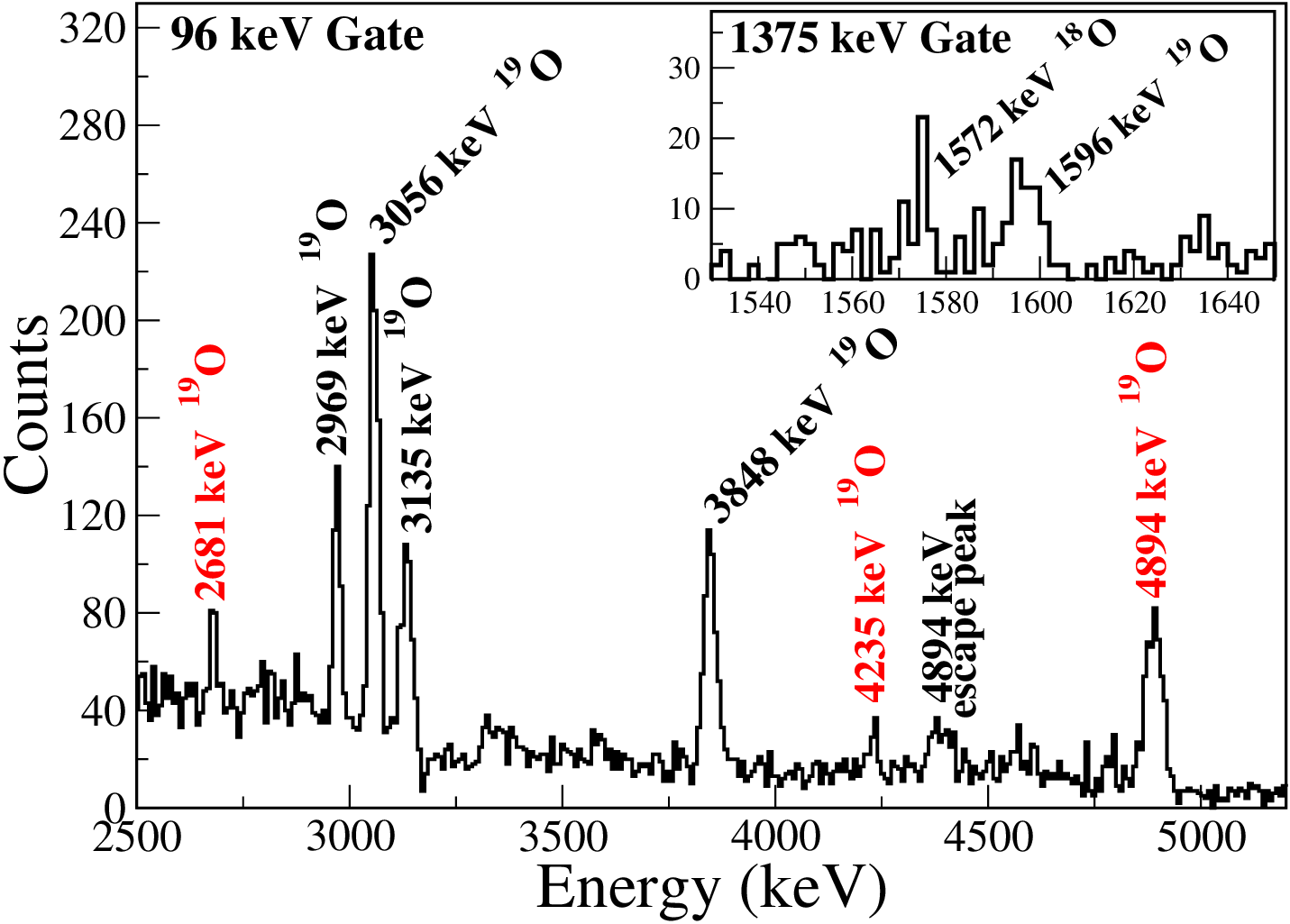}}
 \caption{\label{fig:96gate} (Color online) Portions of the spectra in coincidence with the 96 and 1375 keV $\gamma$ rays. Decays from two of the unbound states are shown (4235 and 4894 keV). Newly observed $\gamma$ lines are colored red.}
\end{figure}

\begin{figure}[!h]
\centerline{\includegraphics[scale=0.7]{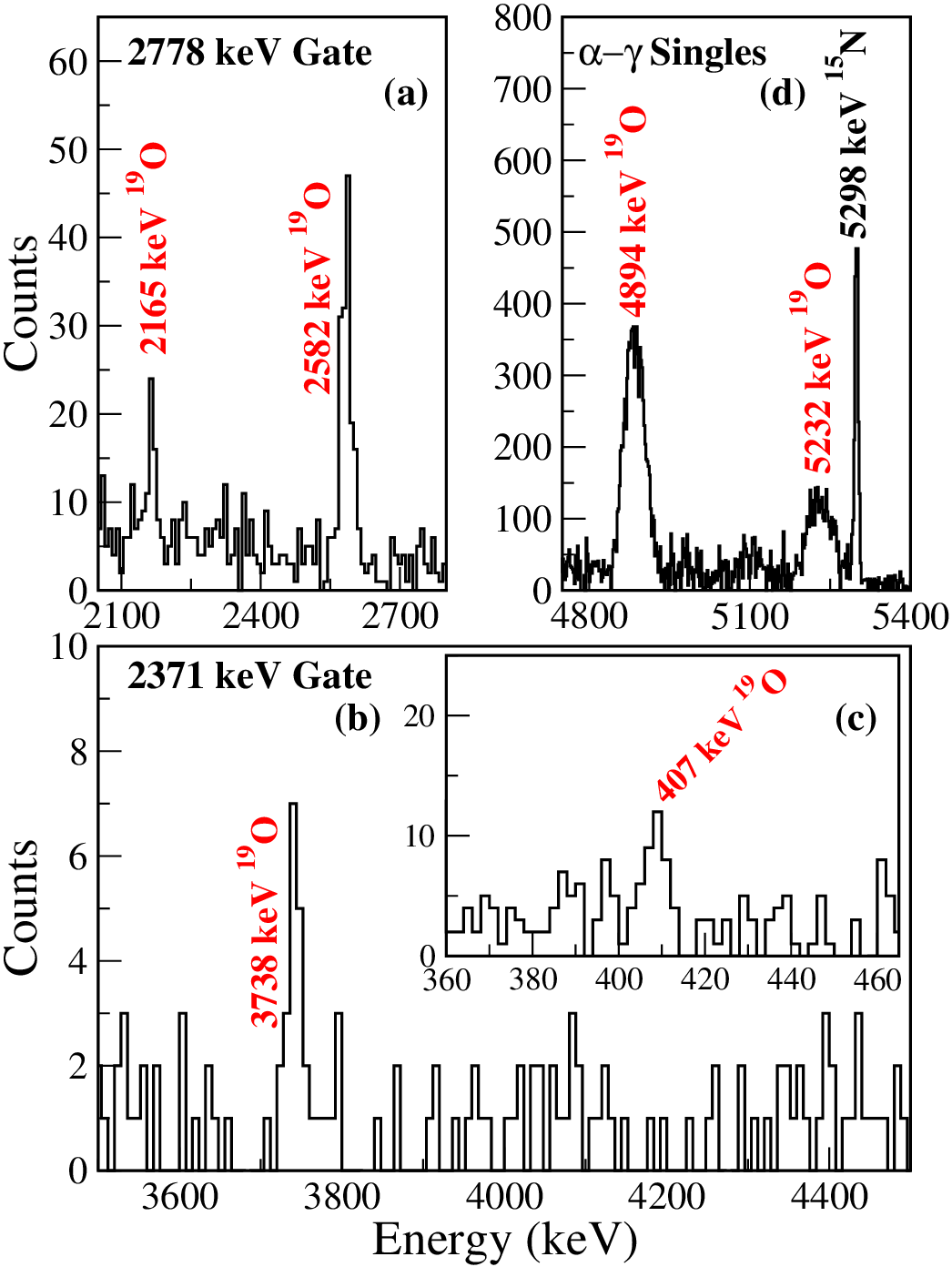}}
 \caption{\label{fig:23712777} (Color online) Portions of the spectra in coincidence with decays from the two highest-spin known lines at 2778 (a) and 2371 keV (b and c) showing three more decay lines from unbound states.  The $\gamma$ singles spectrum (d) gated only by $\alpha$ decays used only 90$^{\circ}$ detectors.  The $^{15}$N line following $\beta$ decay from a contaminant reaction provided a high-energy $\gamma$ energy calibration point. As before, newly observed $\gamma$ lines are colored red.}  
\end{figure}

This was the first study of $^{19}$O using a heavy-ion fusion reaction which favors population of higher-spin states and employs a modern $\gamma$ detector array.  The $\alpha$-$\gamma$ and mostly $\alpha$-$\gamma$-$\gamma$ coincidences were analyzed to build upon the previously known level and decay scheme.  Two of the $\gamma$ decay lines from unbound states can be seen in Figure \ref{fig:96gate} which shows the $\gamma$ spectrum in coincidence with $\alpha$ particles and decays of the 96 keV 3/2$^+$ state.  Gamma decays of the other unbound states can be seen in Figure~\ref{fig:23712777}, especially in coincidence with the 2371 and 2778 keV decays from the states having the highest previously known spin assignments of 9/2$^+$ and 7/2$^+$.  Their placement in the level scheme (Figure \ref{fig:levelscheme}) has been confirmed by reverse gating.

The present experiment has been able to clarify a few open questions about the bound states to which the unbound states decay.  Conflicting $\gamma$ decays of a state at about 3067 keV was pointed out in Ref.~\cite{warburton88}:
only the 1596 keV one was reported in Ref.\cite{broude71} and only the 2969 keV in Ref.\cite{hibou71}. The second decay is not included in the NNDC compilation~\cite{nndc}.  The most obvious resolution is the existence of two different excited states near the same energy.  The present experiment is the first to report both decay lines in the same experiment (see Figure \ref{fig:96gate}), confirming that both $\gamma$ lines exist in $^{19}$O, although this does not distinguish between the possibilities of two different states at $3064\,(2)$ and $3067\,(1)$ keV or one state at about 3065 keV with two decay branches.  The earlier results of different decay modes in different reactions strongly support the doublet possibility with J$^{\pi}$= (5/2$^{-}$) and (3/2$^{+}$) respectively, as has been adopted in Figure \ref{fig:levelscheme}.  The doublet hypothesis is also consistent with shell model predictions but not conclusively proved by this experiment.

The 96-2681 keV coincidence relation establishes a new decay branch of the 2778 keV state to the lowest 3/2$^+$ level, providing further evidence against an older 9/2$^+$ assignment~\cite{moreh64I} and supporting the newer 7/2$^+$ one~\cite{crozier72}.

The observation of allowed $\beta$$^{-}$ decay from the 1/2$^{-}$ g.s. of $^{19}$N to the 3231 and 3945 keV states establishes J$^{\pi}$ = 1/2$^{-}$ or 3/2$^{-}$ to them~\cite{dufour86,sumithrarachchi06}. This confirms the J$^\pi$ = 3/2$^-$ assignment to 3945 keV from earlier  (d,p) reactions~\cite{sen74}.  We see the same $\gamma$ decays from these two states as are reported in~\cite{sumithrarachchi06} and no sign of the 709.2 keV line reported in~\cite{dufour86}, and interpreted~\cite{warburton88} as a 3945 keV $\rightarrow$ 3231 keV transition.  J$^\pi$ = 1/2$^-$ was also reported for the 3231 keV state in a thermal neutron capture experiment ~\cite{nagai07}.  There is also no evidence in the present work for a doublet of states near 3945 keV as suggested in~\cite{fortune74}.

Because the intensities of the $\gamma$ lines from the unbound $4330\,(3)$, $4943\,(4)$, $4990\,(4)$, $5232\,(5)$, $5360\,(3)$, and $6109\,(4)$ keV levels are comparable to the decays from the bound ones, their decays must be predominantly radiative. Five of the six $\gamma$ decays come from levels which correspond in energy within uncertainties to some of the states reported in a comprehensive study of the $^{13}$C($^7$Li,p) reaction \cite{oertzen10}.  However, widths above 4 keV without uncertainties are quoted for all states above S$_n$ in that paper, including values of 5 to 13 keV for the five states observed in the present work to $\gamma$ decay.  Such widths are not consistent with the eV level widths of the observed $\gamma$ decays (see Table \ref{tab:widths}).  Another example independent of the present work is a width of 19 keV listed in Ref. \cite{oertzen10} for the 5151 keV level, while the direct neutron resonance measurement \cite{koehler} reports a width of only 2 keV.  It appears that some of the smaller widths reported in Ref. \cite{oertzen10} overestimate the actual values.

\begin{figure} [!h]
\centerline{\includegraphics[scale=0.7]{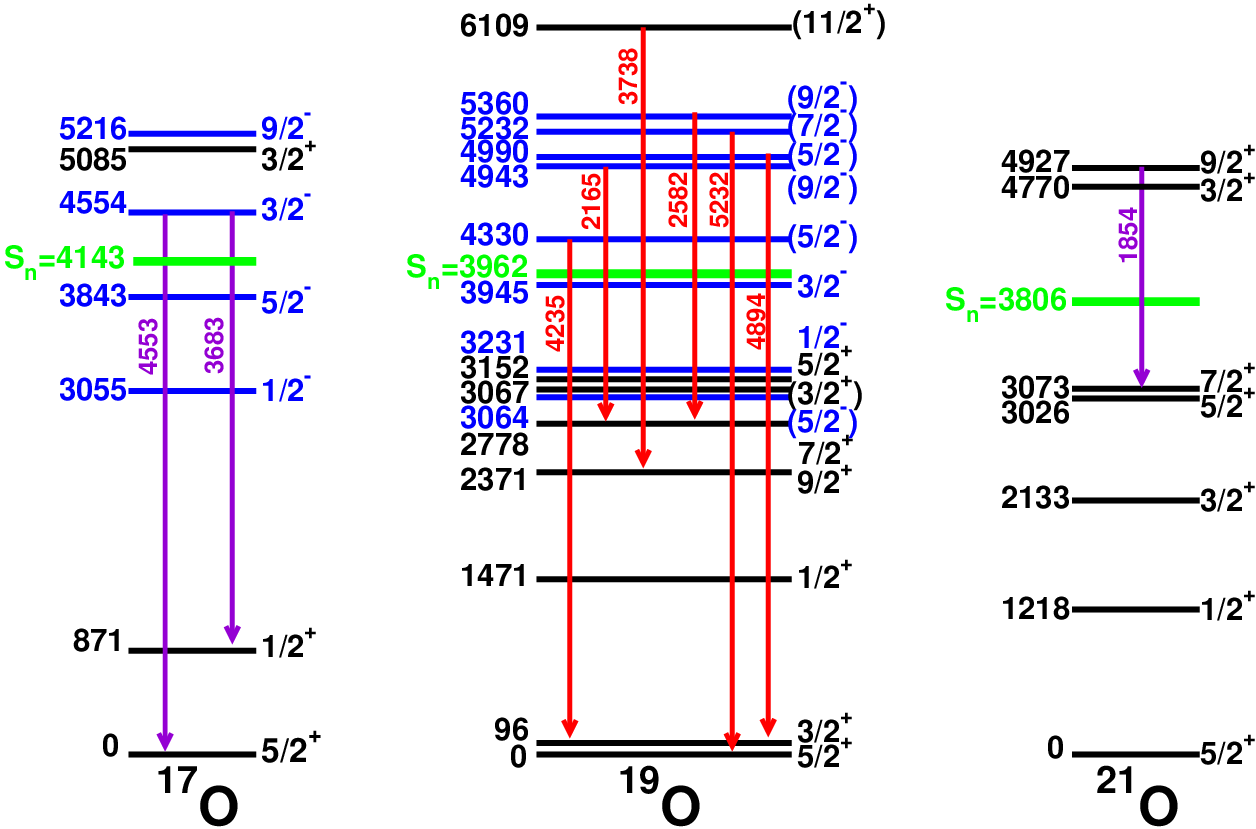}}
\caption{(Color online) A comparison of the experimental level schemes of $^{17,19,21}$O. For $^{19}$O the J$^{\pi}$ values shown in parenthesis are only tentative, as discussed in the text.  The only $\gamma$ decay lines shown are those observed from unbound levels~\cite{igashira,21Odecay}. The two decays in $^{17}$O are very weak branches compared to the neutron decay, whereas the decays in $^{19,21}$O are dominant or 100\% branches.}
\label{fig:oxygencomparison}
\end{figure}

It is informative to compare the nearby odd-A Oxygen isotopes, as shown in Figure \ref{fig:oxygencomparison}. Two $\gamma$ decays have been reported \cite{igashira} from the 4554 keV 3/2$^-$ state in $^{17}$O, which is unbound to $\ell = 1$ neutron decay by 411 keV, but their widths of 1.8 eV each represent only a total radiative branch of about 10$^{-4}$ of the neutron decay width of 40 keV.  These $\gamma$ decays could only be seen in a sensitive neutron capture  experiment and do not correspond to the dominant $\gamma$ decays reported here for $^{19}$O.

More similar to the $^{19}$O decays is a dominant $\gamma$ decay branch from the 9/2$^+$ level in $^{21}$O which is unbound to neutron decay by 1.1 MeV \cite{21Odecay}. The authors calculated a partial lifetime for M1 $\gamma$ decay of 57 fs, but 0.0020 fs for $\ell = 4$ neutron decay with $S$=1.  At least a 50\% $\gamma$ branch would imply a very small neutron spectroscopic factor, less than $3.5\times10^{-5}$.  Their estimated $S$ value from the g$_{9/2}$ occupancy in a shell model calculation is $2\times10^{-4}$.  

Shell model calculations in the \emph{0p-0d-1s} model space using the $\emph {psdu}$ interaction~\cite{utsuno11} which allows full freedom of occupancy in the model space are shown in Figure \ref{fig:levelscheme} for comparison with experiment~\cite{volyacosmo}.  There is reasonable agreement with the states below 4 MeV.  The calculated states above this are dominated by negative-parity ones due to the extra degree of freedom and spin allowed by predominantly one particle-one hole configurations.  It is among these that the best candidates can be found for the radiatively decaying unbound states.  The spin suggestions are shown in parenthesis in Figure \ref{fig:oxygencomparison} and in Table \ref {tab:widths}. The states have been matched based on spin restrictions imposed by the observed $\gamma$ decays, closeness in energy, strongest E-M decays, and highest angular momentum barriers to neutron decay. However, these are only intended as a kind of existence theorem that possible $\gamma$ decaying states exist in the shell model calculations. Some of the spin parity suggestions differ from those of previous works~\cite{fortune77,oertzen10}. However, the previous spin parity suggestions were based solely on empirical methods such as total cross sections and $2J+1$ analysis in Ref.~\cite{fortune77}  and cross section and energy systematics in Ref.~\cite{oertzen10} and hence are tentative.

Decay properties of unbound states in $^{19}$O from the present experiment and Ref.~\cite{koehler} below 6.3 MeV are compared in Table \ref {tab:widths} along with possible correspondences with \emph{pdsu} shell model states.  For the $\gamma$-decaying (neutron-decaying) states, the shell model $\gamma$ decay widths (measured neutron decay widths) are compared with the angular-momentum-limited neutron decay widths assuming perfect overlap between the parent and daughter configurations ($S$ = 1).  The ratio,$S$, of these quantities, listed in the last column, shows the factor by which the neutron decay must be inhibited by the overlap between the initial and final states.  Note that for the $\gamma$ decaying states, $S$ is an upper limit assuming equal $\Gamma_\gamma$ and $\Gamma_n$.  Since $\Gamma_n$ has not been observed, it may well be much smaller than $\Gamma_\gamma$.  All of the spectroscopic factors are well below unity.  Those for the neutron-decaying states span 0.016 to 0.27, while the upper limits for the $\gamma$-decaying states are generally much smaller, ranging from $2.8\times10^{-5}$ to 0.055.  Thus, the experimental observation of radiative decay directly implies very low spectroscopic factors $S$. The \emph{psdu} model space does not include
any of the orbitals \emph{$f_{7/2}$}, \emph{$f_{5/2}$}, \emph{$h_{9/2}$}, nor \emph{$i_{11/2}$}
needed to calculate spectroscopic factors $S$ directly as pure single-particle transitions for the $\gamma$ decaying states. It is not clear that the uncertainties in shell model interactions covering such a wide model space would justify a calculation.

\begin{figure} [!h]
%\centerline{\includegraphics[width=7cm,height=\textheight,keepaspectratio]{figures/nresonance.eps}}
\centerline{\includegraphics[scale=0.7]{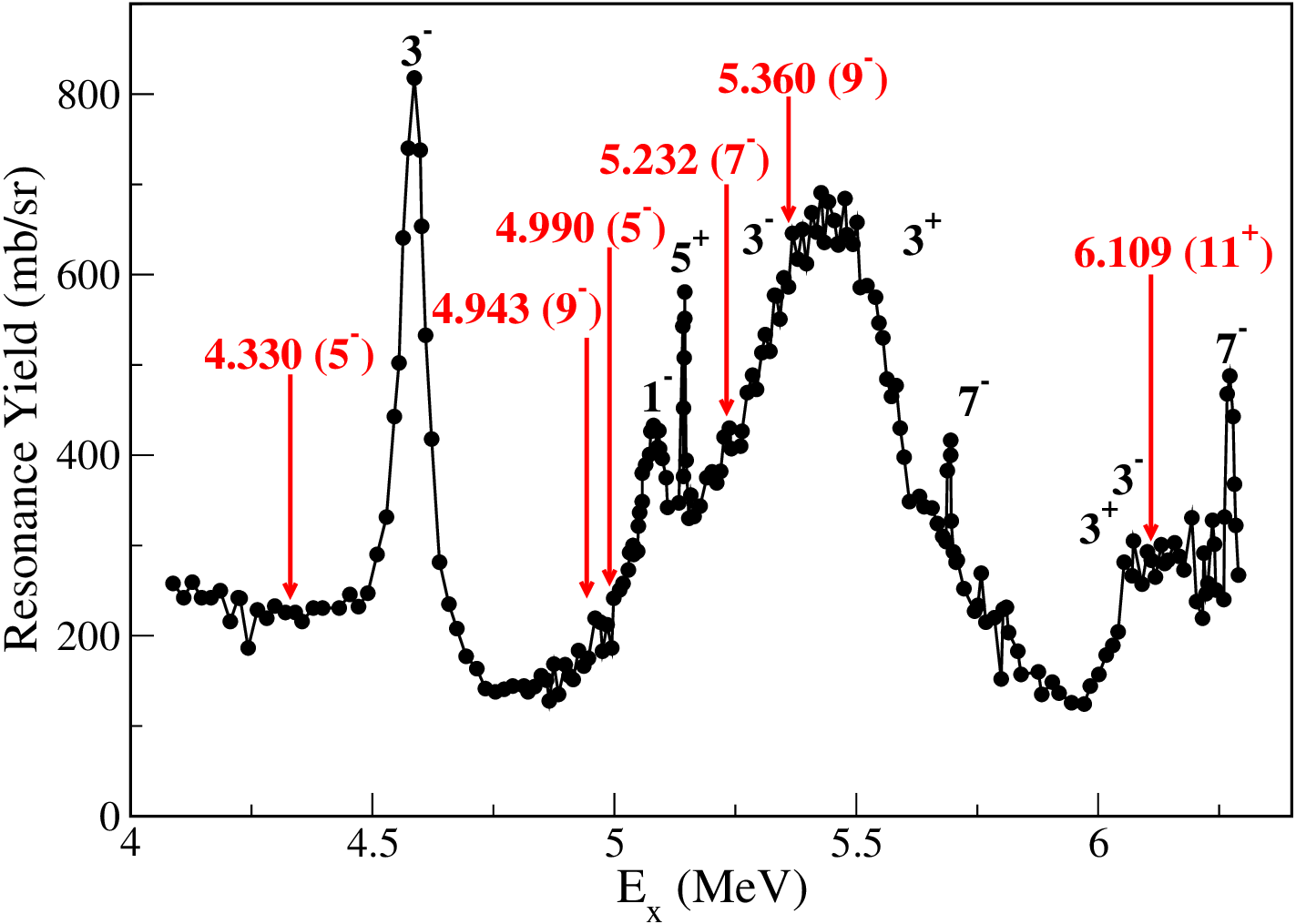}}
\caption{(Color online) A neutron resonance curve on an $^{18}$O target reproduced from~\cite{koehler} with the positions of the $\gamma$ decaying states observed in the present work indicated by red arrows.  Spins are listed as 2J and the ones in red should only be considered as possible spins.}
\label{fig:nresonance}
\end{figure}

Table \ref {tab:widths} and Figure \ref{fig:nresonance} show that the $\gamma$ decaying states are distributed among the neutron decaying ones.  Although the $\gamma$-decaying states are generally higher in spin or lower in energy, this can not be the whole story.  An example is the $7/2^{-}$ states, the lowest of which  decays radiatively while the higher two decay by neutron emission.  The lowest one must also have a small spectroscopic factor (reduced overlap with the  $^{18}$O ground state plus a neutron).  The general idea of aligning experimental and theoretical states in increasing energy order works well for the 7/2$^{-}$ states. However, the predicted bound 9/2$_1^-$ state has not been matched with any experimental states because the predicted  $\gamma$ decays are too weak. Instead the 9/2$_2^-$ and 9/2$_3^-$ states have been matched with unbound $\gamma$ decaying states.  However, one should keep these correspondences in mind as more of a possibility, rather than certain. 

Another characteristic of the $\gamma$ decaying states is that most may be identified with negative parity ones involving an odd number of intruder particles.  This is not surprising since neutron hole states can have higher spins and less overlap with the ground state of $^{18}$O which is unlikely to have substantial intruder components. In this case $\gamma$ decay can occur by parity-changing E1 transitions which are stronger in lighter nuclei. 

$^{19}$O is unique among the Oxygen isotopes in having a higher level density with a reasonably high S$_n$ (see Figure \ref{fig:oxygencomparison}) giving more higher spin states close enough to S$_n$ to allow their $\gamma$ decay to dominate over the usually expected neutron decay. It would be very interesting to compare $\Gamma_n$ and $\Gamma_\gamma$ from the same state, but neutron decay seems to be more of an either-or case than charged-particle decay with its Coulomb barrier.  Two orders of magnitude separate the weakest neutron peaks that can be seen in current resonance studies from the strongest $\gamma$ widths. In fact, there are no clear neutron resonance peaks at any of the energies indicated in Figure \ref{fig:nresonance}.  By comparison the narrow $\Gamma_n$ = 2 keV 5/2$^+$ peak is very visible.

In any event, the present observation of multiple $\gamma$ decaying states above S$_n$ suggests that focused measurements in other nuclei could be fruitful.  The observation of electromagnetic decay competing successfully with the normally much more probable neutron decay provides  significant constraints on the neutron decay width and highlights the exotic structure of these states.  It demonstrates that many exotic unbound states can be missed in neutron resonance measurements so $\gamma$ decay measurements are essential for a more complete spectroscopy. 

\begin{acknowledgements}
This work was supported in part by U.S. National Science Foundation Grants No. 1064819 and No. 1401574 and by the U.S. Department of Energy Grant No. DE-SC0009883. 
\end{acknowledgements}

%\clearpage
\bibliography{19O_bib}% Produces the bibliography via BibTeX.

\clearpage

\clearpage

\begin{table}
\caption{\label{tab:widths}(Color online) Possible \emph{psdu} shell model correspondences for experimentally measured unbound states in $^{19}$O along with uncertainty principle widths.  $\gamma$ (neutron) decaying states are shown in red (black). $\Gamma_\gamma$ is the M1, E1, or E2 width calculated for the observed transition from the shell model calculation with the \emph{psdu} interaction for the $\gamma$-decaying states.  Measured~\cite{koehler} $\Gamma_n$ widths are shown in this column where neutron decay has been observed.  $\Gamma_n (S = 1)$ is the neutron decay width calculated with a Woods-Saxon or square well potential~\cite{volyacosmo}.  $S$ is the neutron decay spectroscopic factor required for equal radiative and neutron decay widths for the $\gamma$-decaying states and the ratio of measured $\Gamma_n$ to penetrability widths for the neutron-decaying states.}
\begin{center}
\begin{tabular}{ccccccc}
\hline
\hline
E$_{x}$(exp)   &    E$_{x}$($psdu$) &  2J$^{\pi}$ &  {\color{red}{$\Gamma _ \gamma$}} or $\Gamma_n$ & $\Gamma _n(S = 1)$ & $S$   \\
 (keV) & (keV) & & (eV)& (eV) & \\
\hline  
\hline 
{\color{red}{4330}}   &  {\color{red}{4433}}  & {\color{red}{5$^-_2$}} &{\color{red}{4.15}} & {\color{red}{214}}  &  {\color{red}{$<0.019$}} \\
4594   &  4847  & 3$^-_2$& 99000 \footnotemark[1] & 363000 &  0.27  \\
{\color{red}{4943}}   &{\color{red}{5405}}  &{\color{red}{9$^-_2$}}& {\color{red}{0.089}} &  {\color{red}{2.1}} &  {\color{red}{$<0.042$}} \\
{\color{red}{4990}}   & {\color{red}{4984}}  & {\color{red}{5$^-_3$}} & {\color{red}{0.20}}  & {\color{red}{7000}} &{\color{red}{ $<2.8\times10^{-5}$}}\\
5073   &  4859  & 1$^-_2$ & 114000 \footnotemark[1] & 713000 &  0.16  \\
5151   &  4518  & 5$^+_3$ & 2000 \footnotemark[1] &  122000 & 0.016  \\
{\color{red}{5232}}   &  {\color{red}{5119}}  & {\color{red}{7$^-_1$}} &{\color{red}{ 5.6}}   & {\color{red}{18400}} & {\color{red}{$<3.0\times10^{-4}$}} \\
{\color{red}{5360}}   & {\color{red}{5745}}  & {\color{red}{9$^-_3$}} & {\color{red}{0.15}}  &  {\color{red}{14}}  & {\color{red}{$<0.011$}} \\
5698   &  5174  & 7$^-_2$ & 2100  \footnotemark[1] & 52000 & 0.040\\
{\color{red}{6109}}   & {\color{red}{7060}}  & {\color{red}{11$^+_1$}} & {\color{red}{0.135}} & {\color{red}{2.45}} & {\color{red}{$<0.055$}} \\
6261   &  5629  & 7$^-_3$ & 12000 \footnotemark[1] \footnotemark[2]& 122000 & 0.10\\
\hline
\hline
\end{tabular}
\end{center}
\footnotetext[1]{measured $\Gamma _n$ from Ref. \cite{koehler}}
\footnotetext[2]{ground state decay branch only}
\end{table}

\end{document}